\documentclass[12pt]{article}

\usepackage{graphicx}
\usepackage[subrefformat=parens,labelformat=parens]{subfig}
\usepackage{amsmath}
\usepackage{amsfonts}
\usepackage{lineno}
\usepackage{multirow}
\usepackage{bibspacing}
\usepackage[super]{nth}
\usepackage{textgreek}
\usepackage{verbatim}

\newcommand\pubnumber{CIPANP2018-Lefebvre}
\newcommand\pubdate{October \nth{1}, 2018}

\def\napoli{Department of Physics\\
McGill University, 3600 rue University, Montr\'{e}al (Qu\'{e}bec), H3A 2T8, CANADA}

\def\Title#1{\begin{center} {\Large #1 } \end{center}}
\def\Author#1{\begin{center}{ \sc #1} \end{center}}
\def\Address#1{\begin{center}{ \it #1} \end{center}}

\newcommand\pubblock{\rightline{\begin{tabular}{l} \pubnumber\\
         \pubdate  \end{tabular}}}
\newenvironment{Abstract}{\begin{quotation}  }{\end{quotation}}
\newenvironment{Presented}{\begin{quotation} \begin{center} 
             PRESENTED AT\end{center}\bigskip 
      \begin{center}\begin{large}}{\end{large}\end{center} \end{quotation}}

\def\beq{\begin{equation}}
\def\eeq#1{\label{#1}\end{equation}}
\def\eeqn{\end{equation}}

\def\beqa{\begin{eqnarray}}
\def\eeqa#1{\label{#1}\end{eqnarray}}
\def\eeqan{\end{eqnarray}}

\let\bar=\overbar

\def\Dslash{\not{\hbox{\kern-4pt $D$}}}
\def\dslash{\not{\hbox{\kern-2pt $\del$}}}

\def\msb{{\bar{\ssstyle M \kern -1pt S}}}

\graphicspath{{figures/}}

\newcommand{\AtlasCopyrightFooter}{%
  \parbox[b]{\linewidth}{%
    \fontseries{m}\fontsize{10}{12}\selectfont
    \copyright\ \the\year \ CERN for the benefit of the ATLAS Collaboration.\newline
    Reproduction of this article or parts of it is allowed as specified in the CC-BY-4.0 license.
  }\par
  }

\begin{document}
\begin{titlepage}

  \pubblock

\vfill
\Title{Muon Spectrometer Phase-I Upgrade for the ATLAS Experiment: the New Small Wheel project}
\vfill
\Author{Benoit Lefebvre\\\small on behalf of the ATLAS Muon Collaboration}
\Address{\napoli}
\vfill
\begin{Abstract}
The instantaneous luminosity of the Large Hadron Collider at CERN will be
increased by up to a factor of five to seven with respect to the design
value. To maintain an excellent detection and background rejection capability in
the forward region of the ATLAS detector, part of the muon detection system will
be upgraded during LHC shutdown periods with the replacement of part of the
present first station in the forward regions with the so-called New Small Wheels
(NSWs). The NSWs will have a diameter of approximately 10~m and will be made of
two detector technologies: Micromegas and small-strip Thin Gap Chambers (sTGC).
The physics motivation for this significant upgrade to the ATLAS detector will
be presented.  The design choices made to address the physics needs will be
discussed.  Finally, the status of the production of the detector modules will
be presented.
\end{Abstract}
\vfill
\begin{Presented}
  \small
Conference on the Intersections of Particle and Nuclear Physics\\
Palm Springs, USA,  May 29 -- June 3, 2018
\end{Presented}
\vfill
\AtlasCopyrightFooter
\end{titlepage}

\section{Introduction}
\label{sec:intro}

The Large Hadron Collider (LHC)~\cite{lhc-paper} is a particle accelerator
located at the CERN laboratory. The initial design luminosity of the LHC is $1
\times 10^{34}$~cm$^{-2}$~s$^{-1}$ with a center-of-mass energy of 14~TeV during
proton-proton collisions. The luminosity has been gradually increased since the
start of LHC operations and reached the design value in 2016. A series of
upgrades are planned over the next decade during Long Shutdown (LS) periods to
further increase the luminosity. In particular, the LHC will be upgraded to its
High-Luminosity configuration (HL-LHC) during LS3, starting in 2024, which will
bring the instantaneous luminosity up to 5 to 7 times the design
value~\cite{hl-lhc}. The LHC is expected to deliver 3000~fb$^{-1}$ of collision
data by its decommissioning in 2037.

An increased LHC luminosity brings more opportunities for physics discoveries to
ATLAS~\cite{atlas-paper}, one of the particle physics experiments of the
LHC. The additional amount of collision data collected will allow for more
precise Standard Model measurements and an increased sensitivity to new physics
processes. Data taking at high luminosity is, however, a challenge for
ATLAS. The planned increase in luminosity will result in increased readout rates
and particle fluences on the ATLAS detector systems. In order to fully benefit
from the physics opportunities of the upgraded LHC, the Phase-I and
Phase-II~\cite{Vankov:2016jyk} upgrades of ATLAS will be carried out during LHC
shutdown periods. In particular, as part of the Phase-I upgrade, the ATLAS muon
identification ability in the forward regions of the detector will be improved.
This upgrade project consists of replacing part of the ATLAS muon spectrometer by
arrangements of muon detector modules called New Small
Wheels~\cite{Kawamoto:1552862} (NSWs) which are targeted for installation
during LS2.

This article is arranged as follows.  A description of the ATLAS muon
spectrometer is given in
Section~\ref{sec:atlas-ms}. Section~\ref{sec:nsw-motivation} presents the
physics motivation behind the NSWs installation. A description of the NSW design
is given in Section~\ref{sec:nsw-design}. A status update of detector
construction is presented in Section~\ref{sec:production}. A short summary of
the article is given in Section~\ref{sec:conclusion}.

\section{ATLAS and the muon spectrometer}
\label{sec:atlas-ms}

ATLAS is a multi-purpose detector of cylindrical geometry located at one of the
LHC interaction points. The ATLAS detector systems are, in order of distance
from the interaction point, the inner detector, the calorimeters and the muon
spectrometer. ATLAS is also equipped with magnet systems which include the
end-cap toroid magnets and the barrel toroid magnet. The magnetic field
generated by the magnet systems bends charged particles such as muons. In the
end-cap regions, muons are typically bent in planes of constant
$\phi$\footnote{ATLAS uses a right-handed coordinate system with its origin at
  the nominal interaction point in the center of the detector and the $z$-axis
  along the beam direction. The $x$-axis points from the interaction point to
  the center of the LHC ring; the $y$-axis points upward. Cylindrical
  coordinates $(r,\phi)$ are used in the transverse plane, where $\phi$ is the
  azimuthal angle around the $z$-axis. The pseudorapidity is defined as $\eta =
  − \ln(\tan (\theta/2))$, where $\theta$ is the polar angle.}. The ATLAS trigger
and data acquisition system (TDAQ) controls the flow of data between the
detector systems and permanent data storage. The TDAQ system follows a
multi-level architecture. The first level, called Level-1, is hardware-based
meaning it uses FPGAs\footnote{Field-Programmable Gate Array} and custom
electronics to quickly process detector hits.

The muon spectrometer is the ATLAS detector system responsible for muon
measurements.  The arrangement of muon detector modules follows an approximate
8-fold azimuthal symmetry. The barrel region has 3 cylindrical shells of
detector modules, or stations. The end-cap regions have 3 disk-shaped
stations. The innermost end-cap station is located at $z=\pm 7.4$~m.

The detectors making up the muon spectrometer are gaseous ionization
chambers. Different detector technologies are used for either Level-1 triggering
or for precision measurements. Thin Gap Chambers (TGC) and Resistive Plate
Chambers (RPC) are used for triggering. Trigger chambers are characterized by a
fast response compared to precision chambers. Hits from trigger chambers are
transmitted to the Level-1 trigger processor. A Level-1 trigger is issued if one
or more muons in the appropriate transverse momentum $p_\text{T}$ range are identified
based on hits from the trigger chambers. Monitored Drift Tubes (MDT) and Cathode
Strip Chambers (CSC) have an excellent spatial resolution and are used for
precision muon track space point measurements. Space points from the precision
chambers are used for the measurement of the muon momentum based on
the Lorentz curvature of the muon tracks in the magnetic field.

\section{Motivation for the New Small Wheel Upgrade}
\label{sec:nsw-motivation}

The ATLAS Level-1 trigger rate increases linearly as a function of the LHC
instantaneous luminosity~\cite{muon-trigger-2015}.  The Level-1 trigger rate
exceeds the ATLAS TDAQ data bandwidth when extrapolating to the luminosity
expected during Run 3\footnote{The ATLAS Level-1 trigger data bandwidth is
  currently 100~kHz and will be 1~MHz after the ATLAS Phase-II
  upgrade.}. Approximately 80\% of Level-1 triggers are associated with single
muon candidates from the end-cap regions~\cite{Kawamoto:1552862}.

\begin{figure}[h]
\centering
\includegraphics[width=0.6\textwidth]{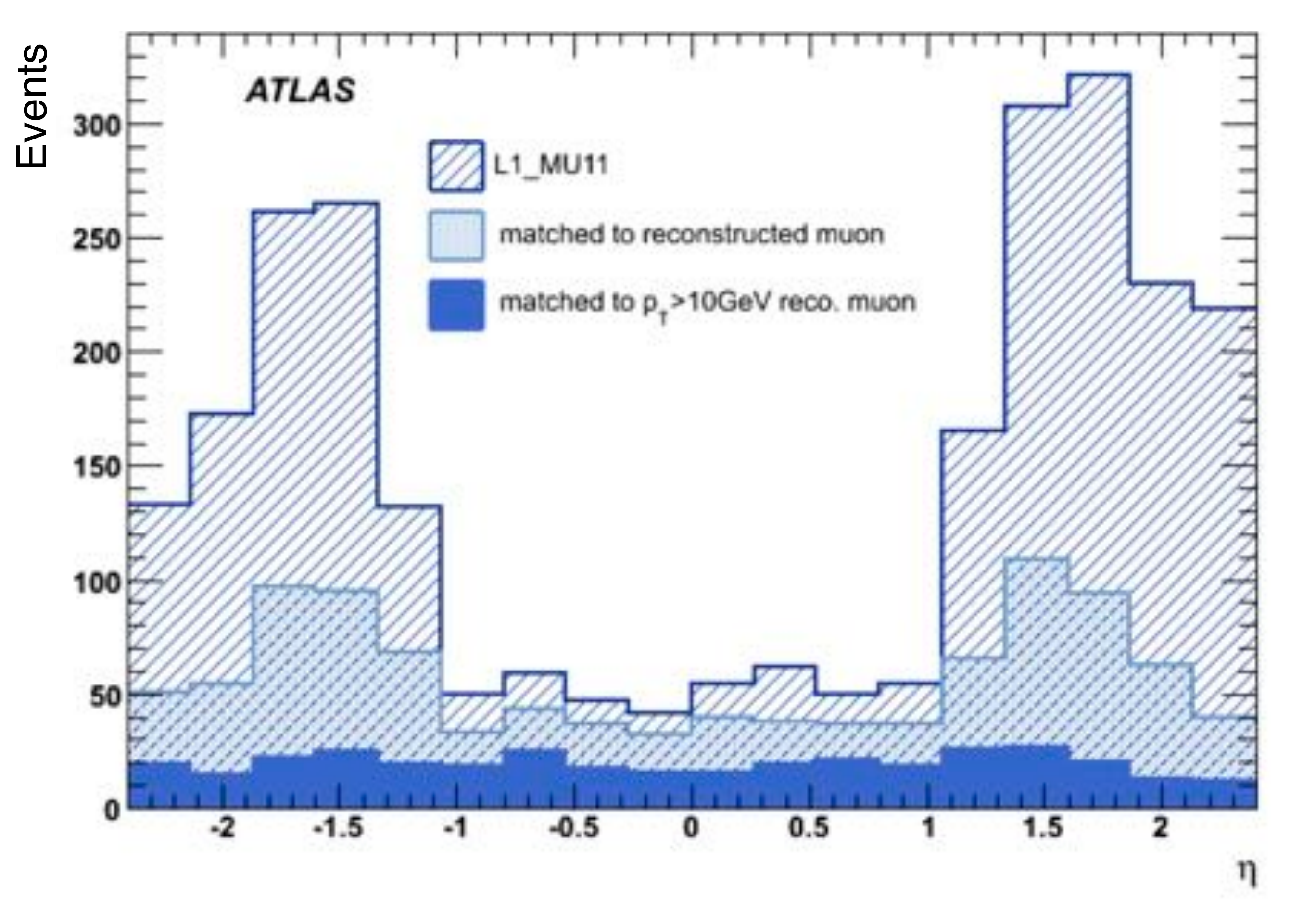}
\caption{Pseudorapidity distribution of muon candidates identified at Level-1
  passing the 11~GeV $p_\text{T}$ threshold (\texttt{L1\_MU11}) and of
  \texttt{L1\_MU11} muon candidates matching a muon reconstructed by an offline
  algorithm, with and without a requirement on its measured
  $p_{\text{T}}$~\cite{Kawamoto:1552862}.}
\label{fig:trigger-rate}
\end{figure}

As shown in Fig.~\ref{fig:trigger-rate}, 90\% of single muon Level-1 triggers
originate from reconstructed track segments that cannot be matched offline with
muons coming from the interaction point. A large fraction of the spurious
Level-1 triggers originates from muon candidates observed in the end-cap
regions. The fake muon rate is explained by background radiation incident on the
end-cap middle station that is incorrectly identified as muons by the trigger processor. The
background radiation consists mostly of neutrons and photons that are generated
in the material between the inner and the middle stations where the end-cap
toroid magnets are located.

A possible workaround for the excessive trigger rate consists of raising the
trigger $p_\text{T}$ threshold from 20~GeV, the nominal value, to 40~GeV or in
applying trigger prescaling. Both options would successfully reduce the Level-1
trigger rate to an acceptable level, but at the expense of reducing the
sensitivity to many physics processes. In particular, a degradation of the muon
trigger efficiency at low $p_\text{T}$ would be detrimental for Higgs studies
because muons are an important signature for many processes involving the Higgs
boson.

Apart from issues arising from the excessive Level-1 trigger rate, a performance
degradation of muon inner end-cap station detectors is expected due to the high
particle fluences. A particle flux reaching up to 15~kHz/cm$^2$ is expected
close to the beam pipe during HL-LHC operations. At that level of particle flux,
the CSC and MDT chambers will suffer from unacceptable
inefficiencies\footnote{Inefficiencies are already sizable but reasonable at the
  nominal LHC luminosity.}.  The loss of muon precision space points will
deteriorate the muon momentum resolution and increase the systematic error of
physics studies.

\begin{figure}[h]
  \center
  \subfloat[]{\label{fig:nsw-algorithm}
    \includegraphics[width=0.50\textwidth]{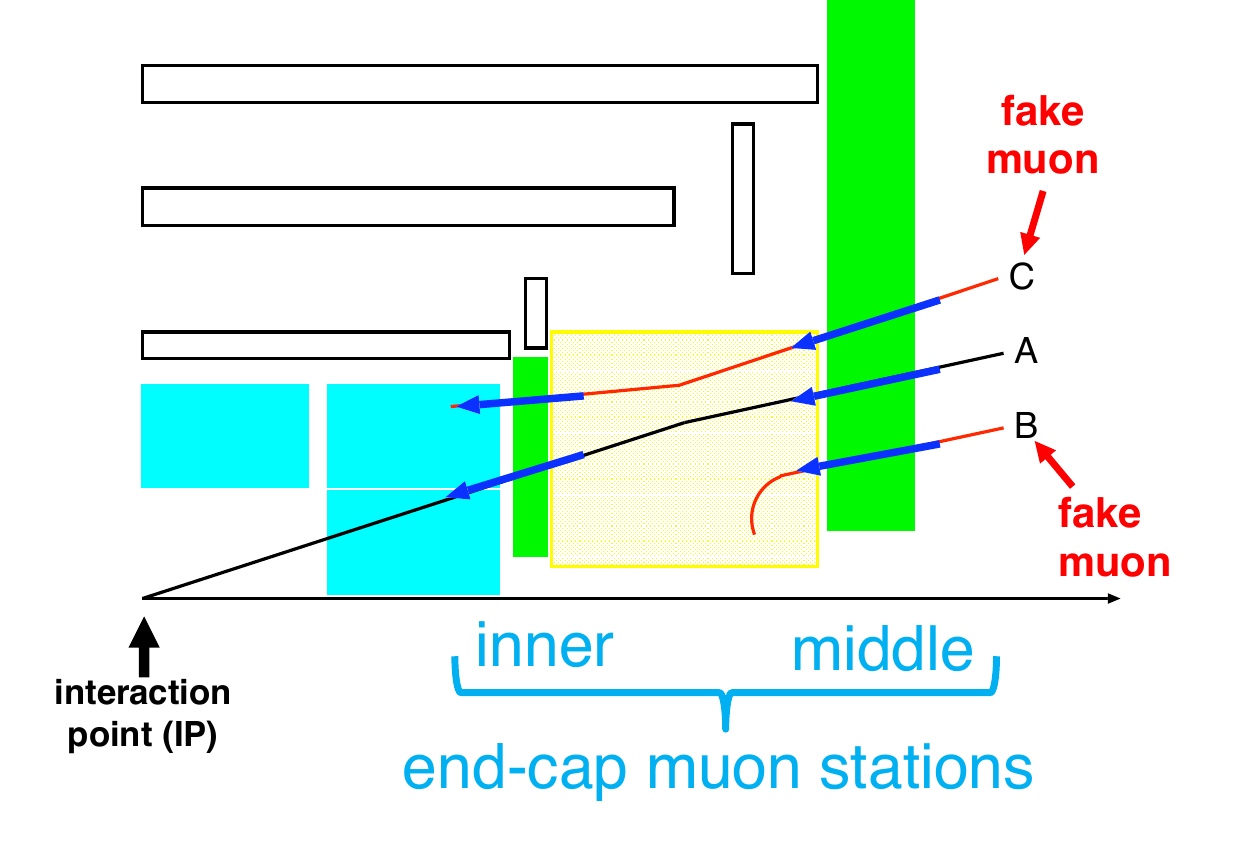}}
  \subfloat[]{\label{fig:nsw:cut-view}
    \includegraphics[width=0.28\textwidth]{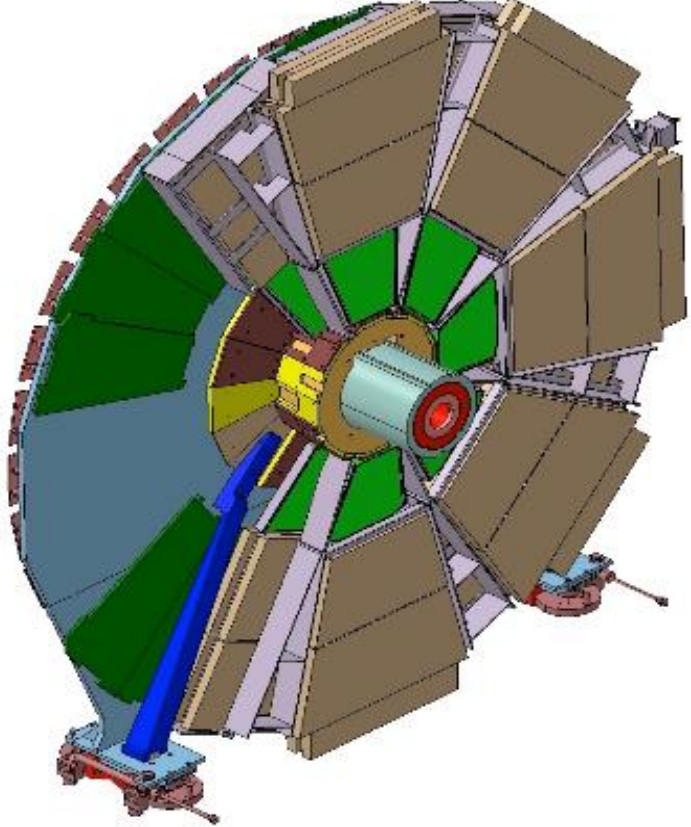}}
  \caption{(a) Schematic diagram of the proposed trigger algorithm used for fake
    muon discrimination viewed on a quadrant cross-section of ATLAS in the
    $r$-$z$ plane.  (b) Cut-away view of a NSW.}
  \label{fig:nsw}
\end{figure}

The proposed solution for reducing the trigger rate is to improve the online
muon identification in order to recognize and reject fake muons. A schematic diagram
of the improved trigger algorithm for muon identification is shown in
Fig.~\subref*{fig:nsw-algorithm} where tracks A, B and C are identified as muons
by the middle station.  In the new trigger scheme, tracks are identified as
muons only if they feature hits in both the inner and middle stations. In
addition, the candidate muon track segment reconstructed by the inner station
must point to the interaction point and match the middle station
measurements. Fake muons (track candidates B and C) typically do not have
these features and would be rejected by the algorithm.

The implementation of this enhanced trigger algorithm necessitates inner station
detectors with excellent online track reconstruction abilities with stable
performances up to the expected level of particle fluences during HL-LHC
operations. The current inner end-cap station detector cannot fullfil these
requirements. Therefore, part of the inner end-cap station will be replaced by
the New Small Wheels (NSWs), shown in Fig.~\subref*{fig:nsw:cut-view} and
described in the next section.

\section{The New Small Wheels}
\label{sec:nsw-design}

The New Small Wheels (NSWs) are disk-shaped arrangements of 8 large and 8 small
pie-slice detector sectors of approximately 10~m in diameter. This geometry was
chosen to match the middle station layout. The detector modules, services and
shielding of the NSWs must fit in the 1110~mm thick envelope left by the current
inner station. The NSW sectors combine the small-strip Thin Gap Chamber
(sTGC)~\cite{MAJEWSKI1983265} and Micromegas\footnote{Micro-mesh gaseous
  structure}~\cite{GIOMATARIS199629} technologies. Both detector technologies
are read out with the VMM ASIC~\cite{vmm3} which performs on-detector peak and
time measurements of the detector signal.

Individual sectors are assemblies of 2 Micromegas wedges placed between 2 sTGC
wedges. Detector wedges are a combination of either 2 Micromegas modules or 3
sTGC modules. Different module types of varying size make up a wedge. The
detector modules are quadruplets, meaning they have 4 independent detector
layers. This multi-layer configuration was chosen for its robustness and
redundancy.

The NSWs are required to have an online angular resolution better than 1~mrad to
match the resolution of the middle station required for the run following the
Phase-II upgrade. In addition, the muon momentum resolution delivered by the
NSWs must be comparable to that of the current inner station aiming for 15\% for
$p_\text{T}$=$1$~TeV muons. Both requirements are satisfied with a spatial
resolution better than 100~\textmu m per detector plane. Finally, the NSWs must
perform a bunch crossing identification of detector hits which requires, to be
achieved, a time jitter better than 25~ns. A number of performance studies, some
of which are described in the following, have proven the ability of the chosen
NSW detector technologies to satisfy these specifications. Specificities of each
detector technology are also described below.

\subsection{sTGC technology}
\label{sec:nsw-design:stgc}

Small-strip Thin Gap Chambers (sTGC) are multiwire chambers that operate in the
quasi-saturated mode. The design of a sTGC is similar to that of the current muon
spectrometer TGC detectors but with an improved spatial resolution and higher
particle flux capabilities. The gas volume of a sTGC is contained between two
segmented resistive cathodes. In the NSW configuration, each gas volume has one
cathode segmented into strips and the other into a pad pattern. Strips are
engraved with a pitch of 3.2~mm and are used for the precise measurement of the
muon trajectory in the bending plane. Pads are used for the online selection of
a specific band of strips to be read out after the passage of a muon thereby
reducing the amount of strip data transmitted to the Level-1 trigger
processor~\cite{guan2016}. The area of pad electrodes varies between
approximately 10 and 500~cm$^2$.

\begin{figure}[h]
  \center
  \subfloat[]{\label{fig:stgc-perf:jitter}
    \includegraphics[height=4.3cm]{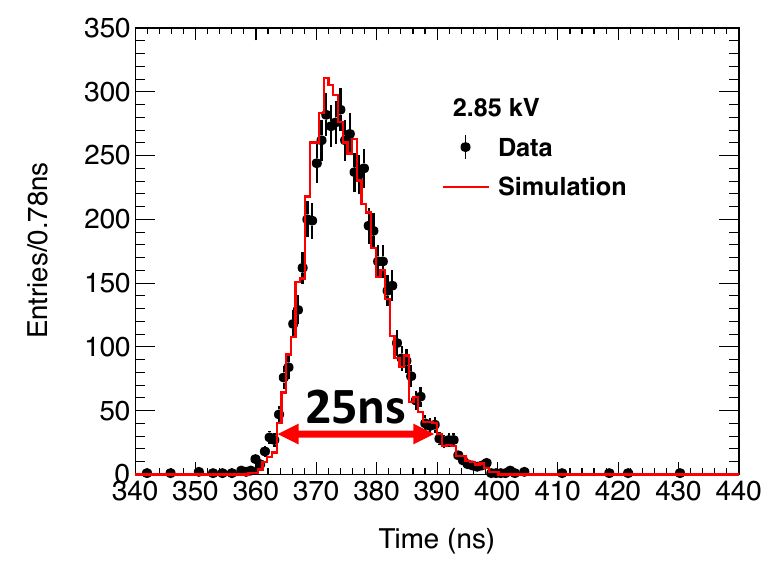}}
  \subfloat[]{\label{fig:stgc-perf:residual}
    \includegraphics[height=4.3cm]{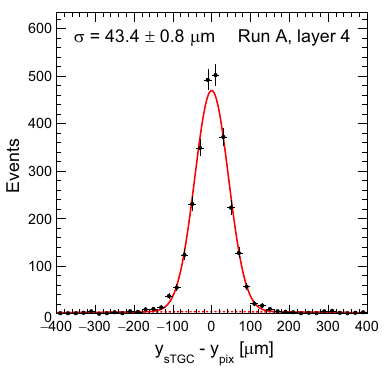}}
  \subfloat[]{\label{fig:stgc-perf:asymmetry}
    \includegraphics[height=4.3cm]{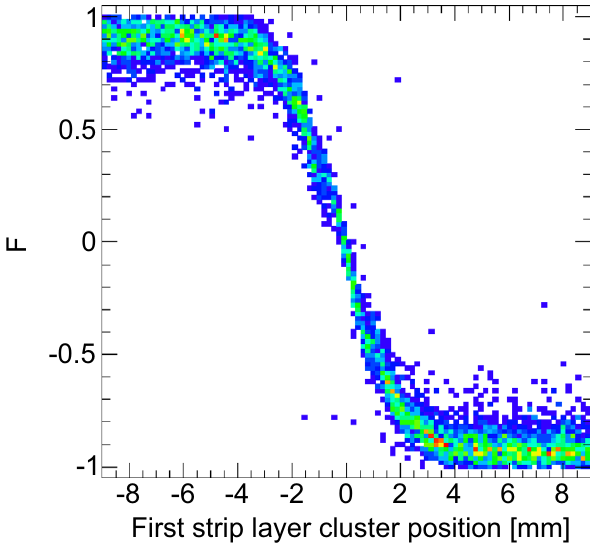}}
  \caption{(a) Measured pad time distribution compared to results from
    simulation~\cite{Kawamoto:1552862}. (b) Difference between sTGC position measurements and a
    precision pixel reference track. The sTGC intrinsic spatial resolution $\sigma$
    corresponds to the width of the distribution~\cite{ABUSLEME201685}. (c) Charge asymmetry between
    neighbour sTGC pads~\cite{ABUSLEME201685}.}
  \label{fig:stgc-perf}
\end{figure}

The time jitter of pads was measured in a beam test at CERN in 2012. The
measured time distribution, shown in Fig.~\subref*{fig:stgc-perf:jitter},
demonstrates that more than 80\% of the hits are within a time window of
25~ns~\cite{Kawamoto:1552862}. The strip spatial resolution and the differential
non-linearity bias were measured in a beam test at Fermilab in
2014~\cite{ABUSLEME201685}. A spatial resolution better than 50~\textmu m was
measured with a perpendicular beam and obtained from the distribution
of position difference between a reference pixel track and the sTGC measurement
shown in Fig.~\subref*{fig:stgc-perf:residual}. The charge sharing between
neighbouring pads was measured in a beam test at CERN
2014~\cite{ABUSLEME201685}. The measured charge asymmetry, as shown in
Fig.~\subref*{fig:stgc-perf:asymmetry}, demonstrates that charge sharing occurs
within a band of 5~mm, which is small compared to the overall dimensions of the
pads.  Integration studies of the final VMM prototype with sTGC detectors are
ongoing at CERN.

\subsection{Micromegas technology}
\label{sec:nsw-design:micromegas}

Micromegas are micro-pattern gaseous detectors with a thick 5~mm drift gap and a
thin 128~\textmu m amplification gap separated by a micro-mesh transparent to
electrons. Micromegas operate with a moderate electric field of 600~V/cm in the
drift region and a strong electric field of 40 to 50~kV/cm in the amplification
region. The ionization electrons created by a muon traversing the drift gap
move towards the micro-mesh, and eventually reach the high-field region where
electron multiplication takes place. The detector signal is picked up by
300~\textmu m wide copper readout strips located opposite to the micro-mesh and
etched on a readout PCB with a pitch of 415~\textmu m. A layer of resistive
strips glued over a thin kapton layer is installed above the readout strips
aiming to improve the stability of the electric field in the amplification
region and reduce the spark probability.

Micromegas quadruplets have two layers equipped with azimuthal strips,
called $\eta$-strips, and two layers equipped with strips arranged in a
stereo-strip configuration at an angle of $\pm$1.5$^\circ$ with respect to the
$\eta$-strips. This strip configuration allows for a precision muon position
measurement in the bending plane and a coarse measurement in the coordinate
perpendicular to the bending plane.

\begin{figure}[h]
  \center
  \subfloat[]{\label{fig:mm-perf:spatial-resolution}
    \includegraphics[height=4.5cm]{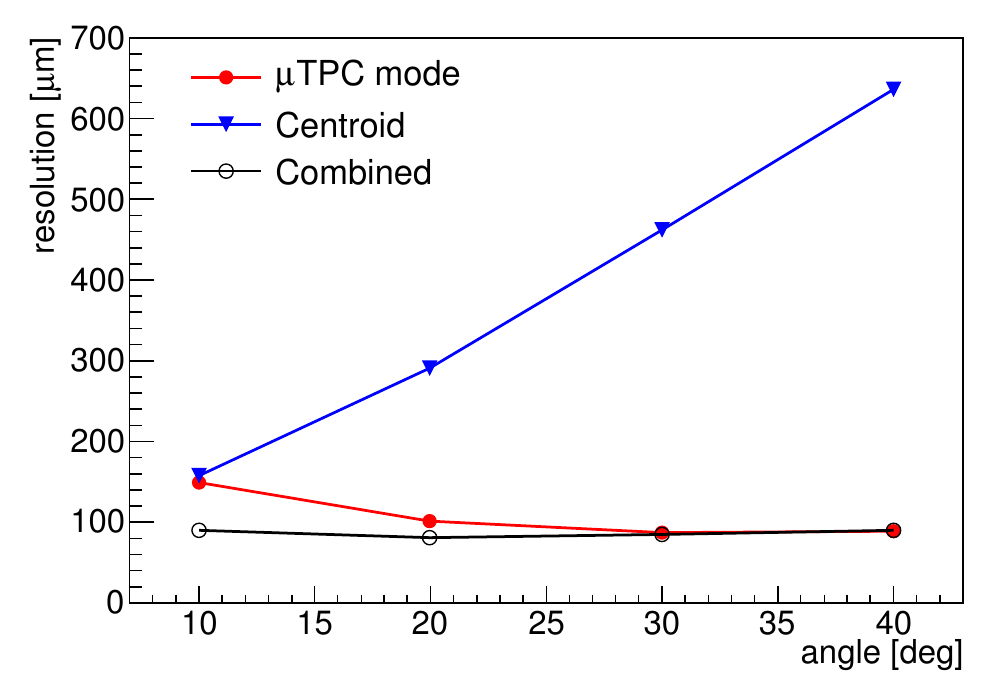}}
  \subfloat[]{\label{fig:mm-perf:timing}
    \includegraphics[height=4.3cm]{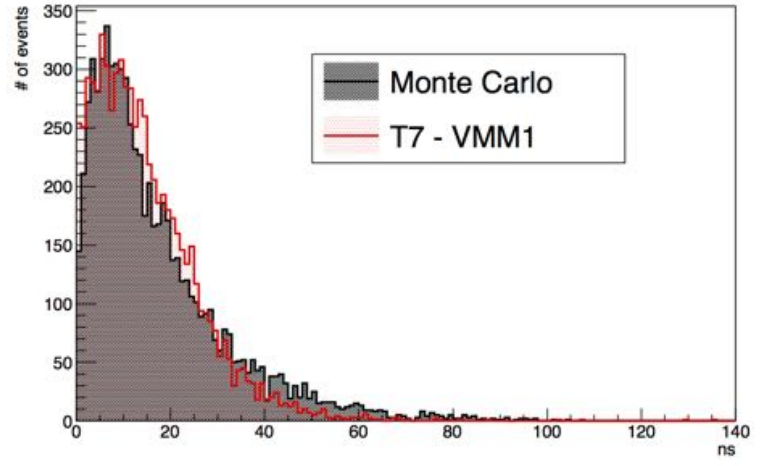}}
  \label{fig:mm-timing}
\caption{(a) Intrinsic spatial resolution of a Micromegas as a function of the
  particle track angle using different analysis
  techniques~\cite{micromegas-tests}. (b) Time distribution of the first strip
  hit compared to results from a Monte Carlo
  simulation~\cite{Kawamoto:1552862}. Both results have been obtained with small
  dimension prototypes.}
\end{figure}

The intrinsic spatial resolution of Micromegas detectors has been measured in
several beam test campaigns from 2012 to the present, first using prototypes of
small dimensions and then on full-size modules.  The spatial resolution as a
function of the track incidence angle obtained for small dimension prototypes is
shown in Fig.~\subref*{fig:mm-perf:spatial-resolution}~\cite{micromegas-tests}.
The spatial resolution obtained when taking the strip charge cluster centroid is
better than 90~\textmu m but degrades as a function of the track angle. The
resolution obtained using the \textmu TPC mode~\cite{micromegas-tests}, which
uses the timing of strip hits, improves as a function of the angle of incidence.
The time distribution of the first strip with a hit, shown in
Fig.~\subref*{fig:mm-perf:timing} was also measured as a way to characterize the
online timing performance~\cite{Kawamoto:1552862}. Most hits are within a time
window of 75~ns.

\section{Detector manufacturing}
\label{sec:production}

Each NSW combines a total of 96 sTGC and 64 Micromegas quadruplets that are
manufactured in member institutes of the ATLAS collaboration.  The production
of detector components such as the sTGC cathode boards and Micromegas readout
boards is carried out in collaboration with industry. The final detector
assembly and testing are realized at CERN.

The NSW performs a precise reconstruction of muon track segments which calls for
a good control of construction non-conformities during detector
manufacturing. In particular, the position of each strip must be known with an
accuracy of 40~\textmu m along the precision coordinate and 80~\textmu m along
the beam to achieve the target NSW momentum resolution.  Stringent tolerances on
the geometry of the readout strips are enforced during construction.  The
alignment between individual strip boards of a quadruplet is achieved using
precision pins. Many steps of the assembly process are carried out on a granite
table to ensure a good planarity of the detector planes.

This section describes specificities of the sTGC and Micromegas
manufacturing. The status of detector production at the time of writing is
given.

\subsection{sTGC manufacturing}
\label{sec:production:stgc}

The sTGC production is divided in 5 productions lines, each responsible for
manufacturing one or two quadruplet types.  In some cases, different steps of
the fabrication and quality control process of a production line are shared
among different institutes\footnote{The sTGC construction sites are PUC (Chile),
  Shandong (China), TRIUMF, Carleton and McGill (Canada), Technion, TAU and
  Weizmann (Israel), and PNPI (Russia).}.

Series production of sTGC modules is well ongoing in the production lines: several
quadruplets have been manufactured and received at CERN. The assembly of the
first sTGC wedge will be finished during Fall 2018. The production of
cathode boards and other detector components is done in parallel to quadruplet
assembly. All sTGC cathode boards will be produced by the end of 2018.

\subsection{Micromegas manufacturing}
\label{sec:production:micromegas}

A total of 5 institutes are in charge of Micromegas quadruplet
production\footnote{The Micromegas construction sites are INFN (Italy), BMBF
  (Germany), Paris-Saclay (France), JINR (Russia) and Thessaloniki
  (Greece).}. Each institute is responsible for the manufacturing of one
quadruplet type. As for the sTGC production, the integration of Micromegas
quadruplets to form NSW wedges is done at CERN.

At this stage, 70\% of the required readout boards have been manufactured and
production will be completed at the beginning of 2019. Series production of
drift and readout panels, the main components of Micromegas quadruplets, is
ongoing. The assembly of quadruplet is started in all construction
sites. First quadruplets have been received at CERN and wedge integration is on
the way to start.

\section{Summary}
\label{sec:conclusion}

The LHC instantaneous luminosity will increase by up to 5 to 7 times the design
value following a series of upgrades planned over the next decade.  In order
to benefit from the physics opportunities offered by an upgraded LHC, the New
Small Wheels (NSWs) will replace part of the muon end-cap station of ATLAS as a
mean to improve the online muon identification capability.

The NSWs will perform the precise reconstruction of candidate muon track
segments and will combine the sTGC and Micromegas detector technologies.
Detector construction for the NSWs is a worldwide effort shared among multiple
physics institutes. Module manufacturing is well ongoing and wedge assembly
has started.

\bibliographystyle{jhep}
\setlength{\bibitemsep}{0.2em}
\bibliography{main}

\end{document}